\documentclass[a4paper]{article}
	
\usepackage{INTERSPEECH2022}
\usepackage{booktabs}
\usepackage{multirow}
\usepackage{array}
\usepackage{lipsum}
\usepackage{float}
\usepackage{url}
\usepackage{makecell}
\usepackage{hyperref}
\hypersetup{
     colorlinks=true,
     citecolor=blue,
}

\title{Investigating Self-supervised Pretraining Frameworks for Pathological Speech Recognition}
\name{Lester Phillip Violeta, Wen-Chin Huang, Tomoki Toda}
\address{
  Nagoya University, Japan}
\email{violeta.lesterphillip@g.sp.m.is.nagoya-u.ac.jp}

\begin{document}

\maketitle
\begin{abstract}

We investigate the performance of self-supervised pretraining frameworks on pathological speech datasets used for automatic speech recognition (ASR). Modern end-to-end models require thousands of hours of data to train well, but only a small number of pathological speech datasets are publicly available. A proven solution to this problem is by first pretraining the model on a huge number of healthy speech datasets and then fine-tuning it on the pathological speech datasets. One new pretraining framework called self-supervised learning (SSL) trains a network using only speech data, providing more flexibility in training data requirements and allowing more speech data to be used in pretraining. We investigate SSL frameworks such as the wav2vec 2.0 and WavLM models using different setups and compare their performance with different supervised pretraining setups, using two types of pathological speech, namely, Japanese electrolaryngeal and English dysarthric. 
Our results show that although SSL has shown success with minimally resourced healthy speech, we do not find this to be the case with pathological speech. The best supervised setup outperforms the best SSL setup by 13.9\% character error rate in electrolaryngeal speech and 16.8\% word error rate in dysarthric speech. 
\end{abstract}

\section{Introduction}

Automatic speech recognition (ASR) is the task of converting speech audio into its text transcripts. Early methodologies trained linguistic, acoustic, and lexicon modules separately \cite{intro1, intro2}; however, up until a decade ago, neural networks showed an improvement of performance while also being significantly easier to train, with all of the modules trained altogether in an end-to-end fashion \cite{e2e_asr1, e2e_asr2}. With the recent successful use of end-to-end models in ASR, many of these have become commercially available as products, bringing in many benefits and conveniences to humans by enabling the control of several devices through speech. With such conveniences, ASR has the potential to benefit many people; however, disabled patients, who would benefit the most from these devices, are hindered from using these owing to their loss of control of their speech organs \cite{pathological_speech}. One common speech pathology is dysarthria, which makes them speak at a slower rate and with a slurred characteristic. Another speech pathology could be due to laryngectomy, which is the replacement of the larynx with an electro-larynx to recreate its function, but results in robotic-like speech production. In both of these speech pathologies, communication becomes difficult, which causes the produced speech to be misunderstood by most commercial ASR devices \cite{inclusive_asr}. Knowing the benefits of ASR, much more research to make ASR accessible to speakers with speech pathologies should be carried out. Unfortunately, only having a small number of publicly available datasets, coupled with the huge data requirements to train networks, have made this a difficult task. 

One mainstream method to resolve minimally resourced ASR is by pretraining, which is training the network on a large dataset of healthy speech before fine-tuning on limited pathological speech datasets. Investigations such as those in \cite{google1} show drastic improvements in performance through a pretraining framework using a Recurrent Neural Network Transducer (RNN-T) pretrained on 1,000 hours of healthy speech data, even with only less than an hour of pathological speech in the fine-tuning stage. Building on this study, increasing pretraining data to roughly 162,000 hours in \cite{google2, google3} makes personalized RNN-T models become capable of outperforming human listeners on speech with low intelligibility rates and allow 90\% of speakers to reach 15\% word error rate even with only using less than 250 utterances in fine-tuning. 
In the aforementioned investigations, using a good pretraining strategy and more data proves that the network can adapt to the pathological speech features easily even when only fine-tuned on a small number of data. However, one limitation in the previously mentioned works is that the model is pretrained in a supervised fashion, which heavily relies on having a dataset that provides both the speech audio and its corresponding text labels to succeed. Although many speech data has been made publicly available, large-scale data labeling is still a very tedious task, making it not a sustainable pretraining method for use in the future.

In this paper, we investigate a novel pretraining method called self-supervised learning, which provides flexibility in the training requirements by removing the need for labeled data to train the network. The general idea of this method is to make the model learn a universal representation of the data using joint embedding objective functions \cite{joint-emb, joint-emb2} to compare embedding representations of the data. For speech in particular, several frameworks have been proposed \cite{yang2021superb} such as generative modeling \cite{generative}, discriminative/contrastive modeling \cite{w2v-2,w2v-xlsr,wavlm}, and multitask learning \cite{multitask}. Research in \cite{analysis_dysasr} has shown contrastive-learning-based SSL pretraining capable of outperforming supervised pretraining on dysarthric speech; however, the research could still be further extended by using more SSL frameworks, while also using other supervised pretraining setups as baseline. 

We contribute the following points in this investigation:
\begin{itemize}
  \item We investigate two self-supervised learning pretraining frameworks (contrastive and masked region prediction) for pathological speech recognition and compare them with supervised pretraining.
  \item We analyze two types of pathological speech, namely, electrolaryngeal and dysarthric.
  \item We show that although SSL pretraining has been successful on minimally resourced healthy speech, extensive experiments show that this is not exactly the case for pathological speech.
\end{itemize}
\begin{figure*}[t!]
  \centering
  \includegraphics[width=15cm]{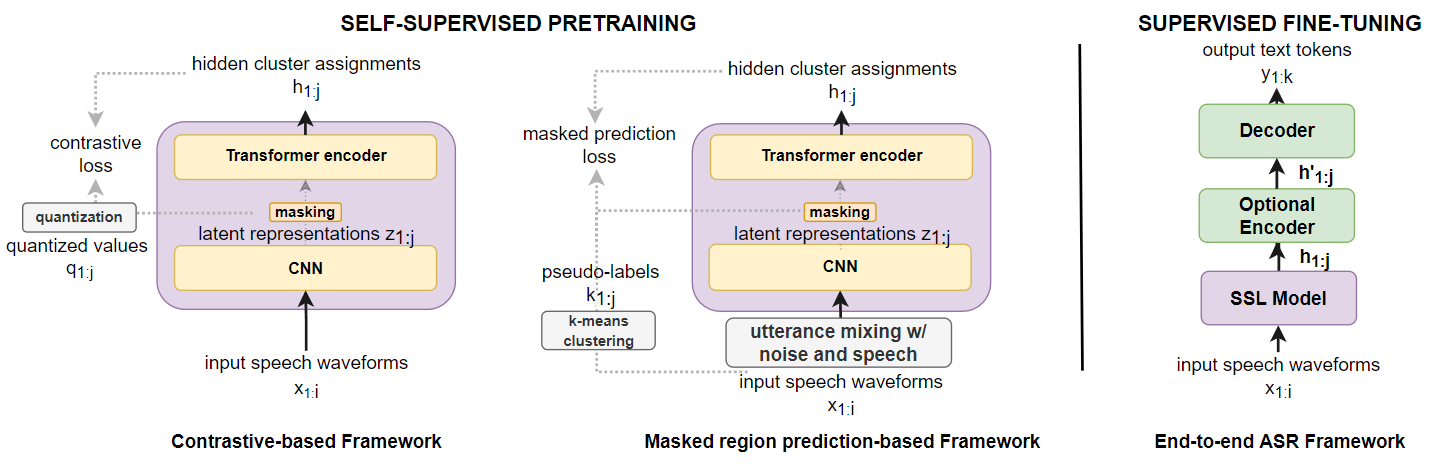}
  \caption{Overview of SSL training. The SSL models wav2vec 2.0 \cite{w2v-2} (left) and WavLM \cite{wavlm} (middle) are first pretrained by a contrastive or masked region prediction framework. The SSL model is then fine-tuned by supervised training, treated similarly to an end-to-end model (right).}
  \label{fig:architecture}
\end{figure*}
\section{Using self-supervised learning pretraining frameworks for ASR}
With the successful use of previous pretraining methods, we explore the effectiveness of a new pretraining framework called self-supervised learning (SSL). 
In this section, we present how SSL models learn speech representations and how they can be used for ASR tasks. An overview of using SSL models, which is divided into pretraining and fine-tuning stages, is shown in Figure~\ref{fig:architecture}.
\subsection{Pretraining SSL models to learn speech representations}

We investigate two self-supervised pretraining frameworks that are similar to the one used in \cite{analysis_dysasr}: contrastive loss \cite{w2v-2} and masked region prediction \cite{wavlm}. The first framework is wav2vec 2.0 \cite{w2v-2}, one of the most widely used SSL frameworks in speech data based on contrastive loss and proven to be successfully used in different minimally resourced speech recognition tasks \cite{ssl_ind, ssl_l2}. 
This framework uses a multilayer convolutional neural network (CNN) to encode speech into a latent space $z_{1:j}$ (Eq. \ref{cnn}). The latent space representations are then quantized as $q_{1:j}$ (Eq. \ref{quantizer}) and randomly masked before being passed onto the multihead self-attention layers of the Transformer encoder backbone (Eq. \ref{tf}). 
The model is optimized using a contrastive loss function by identifying a $q$ from $q_{1:j}$ among a set of distractors $q'$ given the masked context representation vector $h_{1:j}$ from the Transformer encoder.
\begin{equation}
    z_{1:j} = \operatorname{CNN}(x_{1:i})
    \label{cnn}
\end{equation}
\begin{equation}
    q_{1:j} = \operatorname{quantization}(z_{1:j})
    \label{quantizer}
\end{equation}
\begin{equation}
    h_{1:j} = \operatorname{Transformer}(\operatorname{masking}(z_{1:j}))
    \label{tf}
\end{equation}
Next, we investigate a new SSL framework called WavLM \cite{wavlm} that instead uses masked region prediction to learn speech representations. WavLM does this through a denoising and prediction framework, mixing utterances by adding interfering speech and noise to the input speech, and making the model predict the original clean speech. Its architecture is similar to that of wav2vec 2.0, where the noisy input speech is passed onto a set of CNN layers to make $z_{1:j}$ and randomly masked before passing it onto the Transformer network, which then predicts the pseudo-labels of the masked regions $h_{1:j}$. The target pseudo-labels $k_{1:j}$ are generated by performing two iterations of k-means clustering: one on the MFCC of the original clean speech $x_{1:i}$ (Eq. \ref{kmeans1}) and another on the latent representations $z_{1:j}$ (Eq. \ref{kmeans2}). Moreover, a gated relative position bias is added to the Transformer backbone to better capture the sequence ordering of the noisy input speech. Training the model through this denoising framework proves to make the model robust to acoustic variations in the speech.
\begin{equation}
    k_{1:j} = \operatorname{k-means}(\operatorname{MFCC}(x_{1:i})) 
    \label{kmeans1}
\end{equation}
\begin{equation}
    k_{1:j} = \operatorname{k-means}(z_{1:j}) 
    \label{kmeans2}
\end{equation}
\subsection{Fine-tuning SSL models for ASR tasks}
After learning the speech representations, the SSL model is then fine-tuned to predict the text tokens by attaching a decoder, making it similar to a regular end-to-end framework, as seen in the rightmost portion of Figure ~\ref{fig:architecture}. Given an input speech sequence of \(x_{1:i}\), the SSL model maps these into a hidden latent space as \(h_{1:j}\) and is decoded into the text \(y_{1:k}\). Another encoder can also be optionally added after the SSL model to further preprocess \(h_{1:j}\) into \(h'_{1:j}\). Frameworks such as the Connectionist Temporal Classification (CTC) \cite{ctc} decode \(h_{1:j}\) by predicting a text token for each frame. Other popular frameworks predict the text using attention (ATTN) with decoders such as the Recurrent Neural Network \cite{las} or the Transformer \cite{transformer}. Combining CTC and attention (CTC-A) \cite{ctc-attn, ctc-attn2} through a weighted sum has also proven to be effective to decode the text.

\section{Pretraining implementation of pathological speech recognition}
In the following subsections, we describe how we implement the pretraining frameworks for this investigation. In general, we first pretrain a model on a large dataset to make it learn a general idea of speech and subsequently fine-tune it on the pathological dataset for adaptation.

\subsection{Self-supervised pretraining setup}
After pretraining the SSL model, we fine-tune it by attaching a decoder to its last layer. Two decoder setups for fine-tuning are compared. The first is using the original ASR setup in wav2vec 2.0 and WavLM, where a linear projection layer is attached and trained by CTC. Next, following the convention in \cite{analysis_dysasr}, we train by CTC-A by using a vanilla 2-layer 1024-unit LSTM decoder, as recommended in the ASR setup in \cite{yang2021superb}. As pathological speech is extremely acoustically different from healthy speech, we also investigate the effectiveness of adding an extra encoder to further pre-process the SSL features before passing it onto the decoder. Two additional encoders are compared with preprocessing the SSL representations, the Conformer \cite{conformer} and Transformer \cite{transformer}, owing to the strong ability of self-attention to learn representations. We use two attention heads and two layer blocks for both encoders. We then refer to using the SSL representations directly as using an Identity encoder. We find better performance when only the CNN layers of the SSL models are frozen during fine-tuning. The rest of the SSL model components, the encoder, and the decoder are fine-tuned.

\subsection{Supervised pretraining baseline setup}
We compare SSL pretraining with the following supervised pretraining setups. We use the Transformer \cite{transformer} and Conformer \cite{conformer} as the encoders, both of which have produced the lowest word error rate when using the Librispeech \cite{librispeech} dataset in supervised training\footnote{\url{https://github.com/syhw/wer_are_we}}. Both encoders have eight attention heads, although the Transformer has 18 layer blocks, while the Conformer only has 12 layer blocks. The decoder for both encoders is a Transformer composed of eight attention heads and eight layer blocks, and trained by CTC-A. The pretraining stage outputs and predicts byte-pair encoding (bpe) tokens \cite{bpe}; however, in the fine-tuning stage, we initialize a new output layer to decode English and Japanese character tokens instead. For UASpeech, we also compare our results with those obtained using end-to-end models found in \cite{cuhk} that predict the character tokens by using a CONV+BLSTM encoder trained using CTC and ATTN with mel-scale filter banks and pitch as inputs.

\section{Experimental setup}
\subsection{Datasets}

To pretrain the supervised models, we use the 960 hours of Librispeech \cite{librispeech}. On the other hand, SSL pretraining varies per model, with wav2vec 2.0 pretrained on 60k hours of speech and WavLM pretrained on 94k hours of speech. In the fine-tuning stage, we use two pathological speech datasets, as summarized in Table \ref{tab:datasets}. We first investigate an in-house recorded dataset of Japanese electrolaryngeal speech, which we refer to as ELSpeech in this paper. The dataset is recorded by a laryngeal speaker by speaking 1000 different sentence utterances both with an electrolarynx and with their normal voice. We use 85\% of the data for the training set, and the rest for the test set. Both the healthy and electrolaryngeal speeches are used in the train and dev sets but only the electrolarygneal speech used in the test set. As the dataset is in Japanese, we use the character error rate (CER) as the evaluation metric.

The next dataset we use is UASpeech \cite{UASpeech}, a dataset containing parallel English word utterances of 15 dysarthric speakers and 13 healthy speakers. We follow the recommendations in \cite{cuhk} and use both the dysarthric and healthy speakers to train the ASR system but only use the dysarthric speakers for the test set, and remove excessive silences at the start and end of each utterance using a GMM-HMM system\footnote{\url{https://github.com/ffxiong/uaspeech}}. The train and dev sets are from the B1 and B3 blocks, whereas the test set is from the B2 block. We follow the convention of the said paper and use the word error rate (WER) as our evaluation metric.

\begin{table}[th]
\centering
    \caption{Duration (in number of hours) of each split used in the experiment. UASpeech results are obtained after silence removal.}
    \footnotesize
    \label{tab:datasets}
\begin{tabular}{cccccc}
\toprule
& \multicolumn{2}{c}{\textbf{Healthy}} & \multicolumn{2}{c}{\textbf{Pathological}} & \\
\cmidrule(lr){2-3}
\cmidrule(lr){4-5}
\textbf{Dataset} & \textbf{Train} & \textbf{Dev} & \textbf{Train} & \textbf{Dev} & \textbf{Test} \\
\midrule
\midrule
UASpeech \cite{UASpeech} & 2.38 & 0.45 & 3.60 & 0.68 & 2.09 \\
ELSpeech & 1.21 & 0.37 & 1.40 & 0.46 & 0.42 \\
\bottomrule
\end{tabular}
\end{table}

\begin{table}[h!]
\centering
    \caption{CER comparison between using different SSL features and raw features in ELSpeech. ** indicates end-to-end models pretrained with 960h Librispeech.}
    \label{tab:el1}
    
    \scriptsize
\begin{tabular}{rllllc}
\toprule
\multicolumn{1}{r}{\textbf{Sys.}} & \multicolumn{1}{l}{\textbf{Features}} & \multicolumn{1}{l}{\textbf{Encoder}} & \multicolumn{1}{l}{\textbf{Decoder}} & \multicolumn{1}{l}{\textbf{Loss}} & \multicolumn{1}{l}{\textbf{CER\%}} \\ 
\midrule
\midrule
1 &   \multirow{2}{*}{\makecell[l]{Raw\\waveform}}                                          & \textbf{Conf.**}   & \textbf{Transf.**}   & \textbf{CTC-A}        & \textbf{27.9}  \\
2 &                                                                                       & Transf.**            & Transf.**   & CTC-A        & 61.5           \\
\midrule
3 & \multirow{4}{*}{\begin{tabular}[c]{@{}l@{}}wav2vec 2.0\\\end{tabular}}          & Identity        & Linear  & CTC         & 93.3           \\
4 &                                                                                       & Identity        & LSTM    & CTC-A       & 89.2           \\
5 &                                                                                       & Conf.              & LSTM    & CTC-A       & 54.1           \\
6 &                                                                                       & Transf.              & LSTM    & CTC-A       & 50.2           \\
\midrule
7 & \multirow{4}{*}{\begin{tabular}[c]{@{}l@{}}wav2vec 2.0\\ (XLSR)\end{tabular}}    & Identity           & Linear  & CTC          & 99.6            \\
8 &                                                                                       & Identity           & LSTM    & CTC-A           & 89.9            \\
9 &                                                                                       & Conf.                 & LSTM    & CTC-A           & 90.2            \\
10 &                                                                                       & Transf.                 & LSTM    & CTC-A           & 71.5            \\
\midrule
11 & \multirow{4}{*}{WavLM}                                                           & Identity          & Linear  & CTC          & 54.9             \\
12 &                                                                                       & \textbf{Identity}  & \textbf{LSTM}    & \textbf{CTC-A}           & \textbf{41.8}    \\
13 &                                                                                       & Conf.                 & LSTM    & CTC-A           & 42.5             \\
14 &                                                                                       & Transf.                 & LSTM    & CTC-A          & 46.0             \\ 
\midrule
\end{tabular}
\end{table}

\subsection{Implementation}
We use ESPnet \cite{espnet, espnet-2020} to implement the end-to-end ASR models and the s3prl frontend \cite{yang2021superb} to use the SSL models. Both toolkits are open-sourced. 
The SSL models we use are the wav2vec 2.0 Large and WavLM Large models, which are both available to the public.  All networks in this investigation target predicting the character tokens. We fine-tune the networks' learning rate in steps of $10^{-n}$ where $n \in \{ 0, 1, 2, 3, 4, 5 \}$ and use the best results from each run. The language models used are based on the Transformer \cite{transformer} architecture and trained on each dataset's training set text. In Tables \ref{tab:el1} and \ref{tab:ua1}, we will refer to the Conformer architecture as "Conf." and the Transformer architecture as "Transf.".

\section{Results}

\begin{table*}[h!]
  \centering
    \caption{WER comparison between using different SSL features, mel-scale filter bank, and raw waveforms in UASpeech. ** indicates models pretrained with 960h Librispeech.}
    \scriptsize
      \label{tab:ua1}
\begin{tabular}{rllllccccc} 
\toprule
& & & & & \multicolumn{5}{c}{\textbf{WER\% of each intelligibility rating}} \\
\cmidrule(lr){6-10}
\textbf{Sys.} & \textbf{Features} & \textbf{Encoder} & \textbf{Decoder} & \textbf{Loss} & \textbf{Overall} & \textbf{Very Low} & \textbf{Low} & \textbf{Mid} & \textbf{High} \\
\midrule
\midrule
1 & \multirow{2}{*}{Mel-scale filter bank}        & Conv+BLSTM \cite{cuhk}             & Linear & CTC       & 58.0 & 87.1 & 62.4 & 55.3 & 37.8 \\
2 &                                     & \textbf{Conv+BLSTM**} \cite{cuhk}  & \textbf{LSTM**} & \textbf{ATTN}      & \textbf{35.0} & \textbf{68.7} & \textbf{39.0} & \textbf{32.5} & \textbf{12.2} \\
\midrule
3 & \multirow{2}{*}{Raw waveform}       & Conf.**                          & Transf.**  & CTC-A     & 67.2 & 90.6 & 73.6 & 64.1 & 46.6 \\
4 &                                     & Transf.**                          & Transf.**  & CTC-A     & 68.2 & 89.6 & 73.9 & 66.2 & 48.7 \\
\midrule
5 & \multirow{4}{*}{wav2vec 2.0}  & Identity                      & Linear & CTC                   & 75.7 & 93.1 & 79.4 & 74.1 & 60.4 \\
6 &                                     & Identity                      & LSTM    & CTC-A               & 96.4 & 97.9 & 97.5 & 95.8 & 95.0 \\
7 &                                     & Conf.                            & LSTM    & CTC-A               & 71.7 & 92.3 & 74.6 & 68.7 & 55.3 \\
8 &                                    & Transf.                            & LSTM    & CTC-A               & 71.8 & 93.8 & 83.7 & 69.7 & 48.4 \\
\midrule
9 & \multirow{4}{*}{WavLM}        & Identity                     & Linear & CTC                   & 91.5 & 97.3 & 92.3 & 90.5 & 87.1 \\
10 &                                     & \textbf{Identity}            & \textbf{LSTM}    & \textbf{CTC-A}               & \textbf{51.8} & \textbf{71.5} & \textbf{50.0} & \textbf{46.0} & \textbf{40.6} \\
11 &                                     & Conf.                           & LSTM     & CTC-A              & 70.1 & 94.1 & 81.3 & 67.1 & 48.5 \\
12 &                                     & Transf.                           & LSTM     & CTC-A              & 77.2 & 94.5 & 83.9 & 75.3 & 60.3 \\
\midrule
\end{tabular}
\end{table*}
\subsection{Comparison of SSL frameworks}
First, we analyze the two SSL model frameworks. As seen in Sys. 12 in Table \ref{tab:el1} and Sys. 10 in Table \ref{tab:ua1}, using the WavLM model with an LSTM decoder outperforms all setups using wav2vec 2.0 in both datasets, as it produces the lowest CER (41.8\%) and WER (51.8\%), making it the best SSL setup. This performance can be attributed to the speech denoising framework used in masked region prediction, which makes the model more robust to acoustic variations in speech features found in pathological speech and the attention-based LSTM decoder, thereby improving the representation learning. A similar trend in both datasets is also seen when using an additional Conformer or Transformer encoder (Sys. 5, 6, 13, and 14 in Table \ref{tab:el1} and Sys. 7, 8, 11, and 12 in Table \ref{tab:ua1}) to preprocess the SSL features, where it degrades the performance with WavLM, but improves wav2vec 2.0. This shows that the wav2vec 2.0 features are not as strong in projecting the pathological speech into a latent space and needs further processing for it to properly work. Moreover, as the SSL models are trained in English and the ELSpeech dataset is in Japanese, we also test the performance of the multilingually pretrained wav2vec XLSR model \cite{w2v-xlsr}, which has been successfully used in decoding languages not included in the pretraining data. However, we do not find any success to this approach, as seen in Table \ref{tab:el1}, where we see a degradation in performance when using XLSR for all setups compared with its wav2vec 2.0 counterparts.


\subsection{Comparison of SSL pretraining and supervised pretraining}
We present two findings when comparing these two pretraining methods. First, we see a trend similar to that in \cite{analysis_dysasr}, where the best SSL setups in Sys. 12 in Table \ref{tab:el1} and Sys. 10 in Table \ref{tab:ua1} outperform the pretrained Transformer setups in Sys. 2 in Table \ref{tab:el1} and Sys. 4 in Table \ref{tab:ua1}, proving that SSL pretraining has the potential to outperform supervised pretraining. 

However, our second finding is that other supervised setups not used in \cite{analysis_dysasr} may still outperform SSL pretraining. As seen Sys. 1 in Table \ref{tab:el1}, using a pretrained Conformer-Transformer model produces the lowest CER (27.9\%) for ELSpeech, outperforming the best SSL setup by 13.9\%. 
On the other hand, although the Conformer-Transformer setup (Sys. 3 in Table \ref{tab:ua1}) did not perform very well in UASpeech, \cite{cuhk} showed that using a pretrained Conv+LSTM model with 40-dimensional mel-scale filter banks as inputs can still outperform our best SSL setup by 16.8\%, as seen in Sys. 2 in Table \ref{tab:ua1}. 

\subsection{Visualizing differences of dysarthric speech in the latent space}
\begin{figure}[th]
  \centering
  \includegraphics[width=8cm]{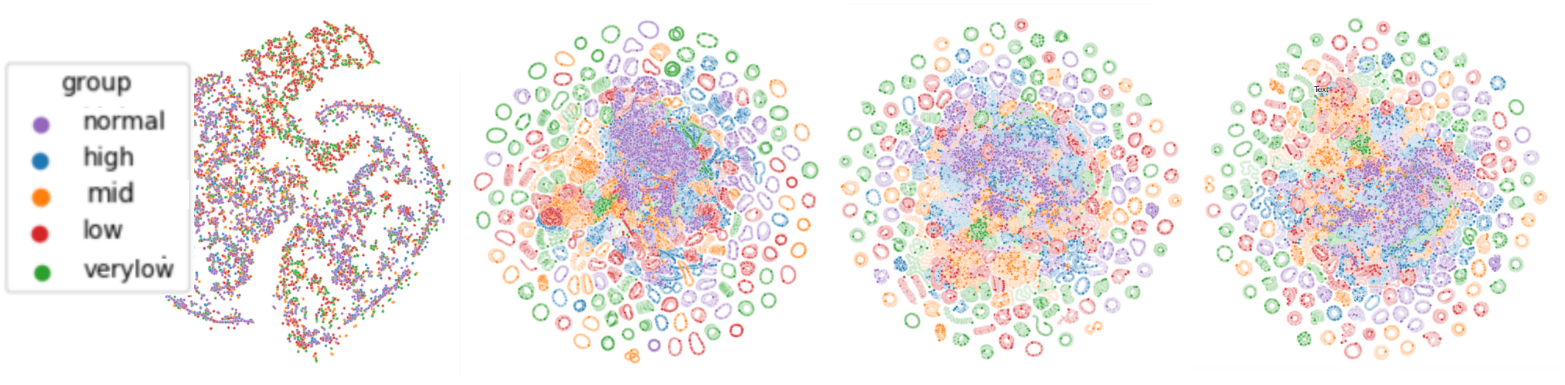} 
  \caption{Pathological speech frame extractions visualized in 2D via t-SNE \cite{t-sne}. Illustrated left to right are mel-scale filter banks with pitch, raw waveforms, WavLM, and wav2vec 2.0} 
  \label{fig:tsne}
\end{figure}

As seen in the results, SSL pretraining did not exactly outperform supervised pretraining in both pathological speech datasets. We attribute this to the stark acoustical differences in pathological speech, as similarly seen in previous research on automatically categorizing dysarthric speech intelligibility \cite{analysis_dyscategorization}, where supervised models outperformed the pretrained SSL model. We then try to visualize the high-dimensional extractions using mel-log filterbank and pitch, raw waveforms, and SSL features with the UASpeech dataset and flatten these on a 2D space using t-SNE \cite{t-sne}. Each point represents one frame in the utterance and is colored according to the speakers' intelligibility rating. We use the first 20 utterances in the B1 block from three speakers in each intelligibility rating. We see in Figure ~\ref{fig:tsne} that raw audio waveforms are projected similarly to the SSL features, proving that there is space to further improve SSL models on pathological speech and improve them at the same level as supervised pretraining. Furthermore, more of the lower intelligibility utterances (green and red) are pushed to the outside, whereas more of the higher intelligibility utterances (purple and blue) are clustered at the center. However, we also see that the mel-scale filter bank extractions seem to blend together the pathological and healthy utterances. One hypothesis we draw from this is that the model may not be able to differentiate the pathological features from the healthy features when using this input, which may have improved the performance. 
\section{Conclusions}
With ASR providing several conveniences to pathological speech patients, more work should be done to make commercial ASR devices useful for them. We investigated two types of SSL-based pretraining frameworks, contrastive and masked region prediction, on two different pathological speech datasets and compared them with a supervised pretraining framework. 
Although SSL pretraining frameworks have shown success with minimally resourced healthy speech, this does not seem to be the case with pathological speech. Further investigations in improving the training strategy can be conducted to improve performance in SSL pretraining. 
Training speaker-dependent models, which has been recommended in several studies \cite{google1, google2, google3}, will be one of the primary targets in the future. 
\section{Acknowledgements}
This work was partly supported by AMED under Grant Number JP21dk0310114, Japan, and JST CREST Grant Number JPMJCR19A3.

\bibliographystyle{IEEEtran}

\bibliography{mybib}


\end{document}